\documentclass[12pt]{iopart}
\usepackage{graphicx}
\usepackage{citesort}

\newcommand{\ket}[1]{|#1\rangle}

\newcommand{\sub}[1]{_{\mbox{\scriptsize #1}}}

\begin{document}

\title[Guided-wave interferometer with mm-scale arm separation]%
{Confinement effects in a guided-wave interferometer with
millimeter-scale arm separation}
\author{J. H. T. Burke, B. Deissler, K. J. Hughes and C. A. Sackett}
\address{University of Virginia, Dept. of Physics, Charlottesville, VA 22904}
\date{\today}

\begin{abstract}
Guided-wave atom interferometers measure interference effects using
atoms held in a confining potential.  In one common 
implementation, the confinement is primarily two-dimensional, and the
atoms move along the nearly free dimension under the influence of an
off-resonant standing wave laser beam.  In this configuration,
residual confinement along the nominally free axis can introduce a phase
gradient to the atoms that limits the arm separation of the
interferometer.  We experimentally investigate this effect in detail,
and show that it can be alleviated by having the atoms undergo a more
symmetric motion in the guide.  This can be achieved by either using
additional laser pulses or by allowing the atoms to freely oscillate
in the potential.  Using these techniques, we demonstrate
interferometer measurement times up to 72~ms and arm separations up to
0.42 mm with a well controlled phase, or times of 0.91~s and
separations of 1.7 mm with an uncontrolled phase.
\end{abstract}

\maketitle

\section{Introduction}
Atom interferometry is a striking example of the particle-wave duality
inherent in quantum mechanics.  Beyond this, however, it has proven to be
a practical and useful technique for several types of precision 
measurements \cite{Berman97}.
Traditional atom interferometers use atoms moving freely through space, with
only occasional manipulation by external fields to control their trajectory.
With the development of laser cooling techniques and Bose-Einstein
condensation, however, it has become possible to implement ``guided-wave''
atom interferometers, in which an applied potential directs the
atomic motion at all times \cite{Shin04,Schumm05,Wang05,Wu05,Garcia06,Jo07}.  
A useful analogy is the comparison between
an optical interferometer constructed using free beams manipulated by
mirrors, versus one using light confined in optical fibres.  The
advantages of guided-wave atom interferometry include the ability to
incorporate more complex trajectories, and also to support the atoms
against gravity over long measurement periods.  It is hoped that
these advantages will eventually allow guided-wave interferometers to exceed
the capabilities of free-space devices.
Guided-wave atom interferometry presents several challenges, however, since it
is critical that the guiding fields impart no
uncontrolled quantum phase to the wave-packets.  

One limitation that is particularly important for interferometers 
using larger arm separation is the effect
of longitudinal variations in the guiding potential.  In several
experiments to date \cite{Wang05,Garcia06,Horikoshi06},
the ``guide'' is really an elongated harmonic trap,
with atomic wave packets moving along the weak axis of the trap.
The nonuniform potential imparts a spatially varying phase shift
to the packets, so that different parts of the packet interfere with
different phases and the overall visibility of the interference is degraded.
In this paper, we explore this phenomenon in detail by experimentally
measuring the phase gradients imposed and confirming their
relation to the visibility.  We also demonstrate that the effect
can be much reduced by using more symmetric trajectories for the
atomic packets.  In this way, packet separations of up to 0.42 mm have
been achieved with controllable overall phase, and
interference is still observed for separations of 1.7 mm. 
In the latter case, however,
environmental noise effects cause the phase to fluctuate randomly 
from one measurement to the next.

One figure of merit for an interferometer is the time integral
of the wave packet separation, which we denote $\chi$.  
This gives directly the accumulated phase for 
measurement of a linear gradient field such as gravity.
To put the results in context,
free space interferometers reach $\chi \approx 500~\mu$m$\cdot$s 
\cite{Peters01,McGuirk02}.  Previous results for our experiment
were limited to $\chi = 6~\mu$m$\cdot$s \cite{Garcia06}, 
and other approaches have yielded $\chi \approx 2~\mu$m$\cdot$s 
\cite{Su07,Jo07}.  The improvements reported here yield
$\chi = 15~\mu$m$\cdot$s with controllable phase, and
850~$\mu$m$\cdot$s with uncontrolled phase.

\begin{figure}
{\includegraphics[width=6in]{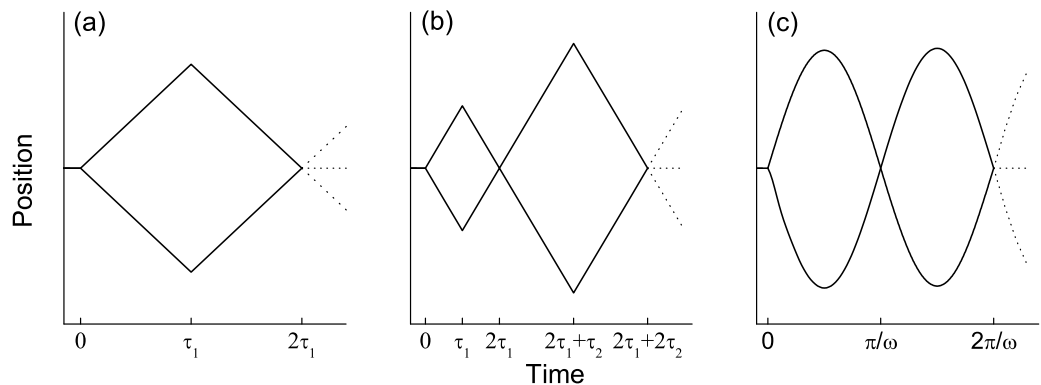}}
\caption{\label{fig:Tra} Wave packet trajectories for various
interferometer configurations.  (a) Single-sided, with leg
duration $\tau_1$.  (b) Double-sided, with leg duration
$\tau_1$ for the first half and $\tau_2$ for the 
second half.  (c) Free oscillation in a waveguide
with harmonic confinement at frequency $\omega$. Note that the
time and position scales vary considerably for the different sub-figures.}
\end{figure}

Our interferometer has been described in detail in
Ref.~\cite{Reeves05,Garcia06}, and the basic operation
is illustrated in Fig.~\ref{fig:Tra}(a). 
A Bose-Einstein condensate of
$3\times10^{4}$ $^{87}$Rb atoms is prepared in the $F=2, m_{F}=2$
hyperfine state and held in a time-orbiting potential waveguide with
harmonic confinement at frequencies $(\omega_x, \omega_y, \omega_z)
\approx 2\pi\times(6.0, 1.1, 3.3)$~Hz. We note that this confinement
is weaker than that used in other comparable experiments. The atoms
are manipulated using an off-resonant laser tuned 12 GHz blue of the
$5S_{1/2} \leftrightarrow 5P_{3/2}$ principle transition at
wavelength $\lambda = 780$~nm. The beam is reflected to form a
standing wave aligned to the $y$-axis of the waveguide.  By applying a
double pulse of this standing wave \cite{Wu05b}, atoms at rest in the $\ket{0}$
momentum state can be driven to a superposition of states moving
with momentum $p = \pm 2\hbar k$: $\ket{0} \rightarrow (\ket{2\hbar
k}+\ket{-2\hbar k})/\sqrt{2}$, where $k = 2\pi/\lambda$ is the
wavenumber of the standing wave laser. In our case, the moving atoms
acquire a speed of $v_0 =  11.7$~mm/s. As described in
Ref.~\cite{Hughes07}, this operation can be made very precise. After
this `splitting' is complete, the packets propagate away from each
other in the $y$ direction for a time $\tau_1$. The standing wave is
then applied a second time, with intensity and duration such that
the `reflection' transition $\ket{2\hbar
k}\leftrightarrow\ket{-2\hbar k}$ is driven. After a total time
$2\tau_1$, the packets return to their initial position. A third laser
pulse, identical to the first splitting pulse, can then be applied.
If there is no phase difference between the two packets of atoms,
this final pulse reverses the action of the first, and all the atoms
come back to rest. However, if we apply a phase $\theta$ between the packets, some portion of the atoms will continue in the moving
states.  
The fraction of atoms in the rest state, $N_0/N$, is
given by $\cos^2(\theta/2)$. In practice, the phase $\theta$ is
controlled by adjusting the
frequency of the standing wave 
laser before the final pulse \cite{Garcia06,Burke05}.  
We refer to the simple trajectory of Fig.~\ref{fig:Tra}(a) as
a single-sided interferometer.

In this geometry, it is not difficult to see how the axial confinement
can produce a phase gradient across the packets.  The packets have
a length along the guide of $L$, set by the size of the initial condensate.
As the packets move apart, the leading edge of the packet experiences
a larger potential energy than the trailing edge, since it is further away
from the bottom of the potential.  This edge therefore develops
a larger phase, which continues to accumulate through the entire trajectory.
Since the leading edge of one packet interferes with the trailing edge
of the other, the phase difference between the packets has a spatial gradient.

This problem was first recognized by 
Olshanii and Dunjko \cite{Olshanii05}, who used it to explain
the limited interferometer performance observed by Wang {\em et al.}
\cite{Wang05}.  Later,
Horikoshi and Nakagawa \cite{Horikoshi06} 
demonstrated an interferometer using a similar
geometry and showed that it too was limited to an arm separation
consistent with that imposed by the phase gradient effect.
In these experiments, interactions between the atomic packets
also played a significant role.  As the packets separate, different
parts of the packets interact for different times. This also induces
a phase gradient, which can be comparable to the confinement effect
described above.  Stickney {\em et al.}\ \cite{Stickney07} have 
suggested that the
confinement and interaction gradients can be made to cancel in some cases,
allowing the arm separation to be extended.

Due to the weak confinement of our waveguide, our experiment
operates in a somewhat different regime, where the effects of
interactions are small and the confinement effects dominate.  We
assert that any interferometer with well-separated arms will
necessarily be in this regime, since the interaction phase stops
growing once the packets are separated while the confinement effect
continues to accelerate.  The solution we examine here is to use a
more symmetric interferometer geometry, as shown in Fig.~\ref{fig:Tra}(b).
Here the packets are not
immediately recombined when they return to the centre.  Instead,
they are allowed to continue propagating for another interval
$\tau_2$, whereupon they are reflected a second time. They finally
overlap again and are recombined at total time $2(\tau_1+\tau_2)$, as
illustrated in Fig~\ref{fig:Tra}(b). In the second
half of this `double-sided' interferometer, the leading and
trailing edges of the packets exchange roles, so that the
confinement phase accumulates with opposite sign.  
When $\tau_1 = \tau_2$, the phases from the two halves nearly cancel
and the arm separation can be
significantly extended. We made use of this type of trajectory
previously \cite{Garcia06}, but here we investigate the phase
cancellation effect in detail.  We also demonstrate an arm
separation about twice that previously achieved, due to various
improvements in the apparatus \cite{Hughes07}. Even larger separations can be
obtained using no reflection pulses at all.  If the packets are
simply allowed to oscillate freely for one period of the harmonic
guide potential as in Fig.~\ref{fig:Tra}(c), 
then the two halves of the cycle are entirely
symmetric and the phase gradient cancels more precisely.  The phase output of this 
interferometer is currently unpredictable, however,
which renders its use impractical.

The remainder of the 
paper is organized as follows: Section 2 describes the theoretical
analysis and experimental investigation of the one-sided `single-reflect'
interferometer.  Section 3 discusses the two-sided interferometer.
Section 4 presents the results of the free oscillation interferometer,
and Section 5 offers a summary and conclusions.

\section{Single-sided interferometer}
We estimate the phase gradient for our experimental conditions using a
simple semi-classical theory.  We neglect any internal dynamics of the
packets themselves, treating them as rigid bodies.  This is justified
since the internal motion occurs on the time scale of the confinement
frequency $\omega$, and the confinement effects constrain
the interferometer time $\tau_1$ to $\omega\tau_1\ll 1$, as will be seen.
Additionally, we assume that the atoms start at rest at the minimum of
the trap.  Non-zero initial motion of the condensate is not difficult
to include, and can change the phase development.  However, in
our experiment, this effect is  small enough to neglect.
The potential along the $y$-axis of the guide taken to be
\begin{equation}
V\sub{ext}(y) = \frac{1}{2} m\omega^2 y^2.
\end{equation}
We define $y_{1}(t)$ and $y_2(t)$ to be the positions
of the centres of the two packets produced by the splitting operation.
Within a packet, the position relative to the centre is labelled
$\xi_1$ or $\xi_2$.

The phase acquired by packet 1 over the total time of the interferometer is
\begin{equation}
\phi_1(\xi_1) = \frac{1}{\hbar}\int_0^{2\tau_1} V\sub{ext}(y_1+\xi_1) dt =
\frac{m\omega^2}{2\hbar} \int_0^{2\tau_1} [y_1(t)+\xi_1]^2 dt.
\label{eq:phi}\end{equation}
Before the reflection pulse, the packet trajectory is given by
\begin{equation}
y_1(t) = \frac{v_0}{\omega}\sin(\omega t)  \quad\quad
(0<t<\tau_1).
\end{equation}
At time $\tau_1$, the reflection pulse provides a velocity kick of
$-2v_0$, so the subsequent trajectory is
\begin{eqnarray}
y_1(t) = \frac{v_0}{\omega}&\sin(\omega\tau_1)\cos[\omega(t-\tau_1)]\nonumber \\
&+\frac{v_0}{\omega}[\cos(\omega\tau_1)-2]\sin[\omega( t-\tau_1)]
\quad (\tau_1<t<2\tau_1).
\end{eqnarray}
The second packet follows a similar trajectory with
$v_0$ replaced by $-v_0$ everywhere.

Notice that
the `reflect' pulse really is a momentum kick and not a true reflection.
This has an effect that will be important in Section 3.
As the packets move to higher potential they slow down, with packet 1
reaching a velocity $v=v_0-\delta v$. The reflection pulse alters this
to $-v_0 -\delta$, so the packet is then moving faster than $v_0$.
As the packet returns to the origin, it accelerates further, to approximately
$v=-v_0-2\delta v$.  Because of this, at time $2\tau_1$ the two packets are
not precisely overlapped, but have moved past one another \cite{Stickney07}.
This can be corrected by modifying the length of time 
between reflect and recombination, 
but in our experiment the effect is negligible for
the single-sided interferometer and we have $y_1(2\tau_1) \approx y_2(2\tau_1)$,
so that $\xi_1 \approx \xi_2 \equiv \xi$.

Evaluation of the integral in (\ref{eq:phi}) gives
\begin{equation}
\phi = \phi_1(\xi)-\phi_2(\xi)
= -4k\xi [\cos(2 \omega \tau_1) -2 \cos(\omega \tau_1) + 1],
\label{eq:SPhi} \end{equation}
using $v_0 = 2\hbar k/m$.
For $\omega\tau_1 \ll 1$, the phase is approximately
$4 k\xi(\omega\tau_1)^2$.  A smaller effect comes from the interaction
between the packets while they are still overlapped.  
To include this, we treat the system
as one-dimensional, so that the initial condensate satisfies the
Gross-Pitaevskii equation
\begin{equation}
i\hbar\frac{\partial}{\partial t}\Psi(y,t)=\left[
-\frac{\hbar^2}{2m}\frac{\partial^2}{\partial^2 y}
+V_{ext}(y)+g_{1D}\left|\psi(y,t)\right|^2\right]\Psi(y,t),
\label{eq:GP}
\end{equation}
where $g_{1D}$ is the one dimensional coupling constant \cite{Olshanii98}.
The actual experiment is not accurately in the one-dimensional limit, but
it provides a convenient approximation for what is anyway a modest correction.
In the Thomas-Fermi limit the number density is
\begin{equation}
n_{1D}(y) = \left\{
\begin{array}{l l}
  \displaystyle{\frac{\mu}{g_{1D}}}
  \left(1-\displaystyle{\frac{y^2}{L^2}}\right) & \quad -L<y<L \\
  \rule{0pt}{4ex} 0 & \quad \mbox{otherwise}\\ \end{array} \right.
  \label{eq:num}
\end{equation}
for chemical potential $\mu$ and condensate half-length
$L=\sqrt{2\mu/m\omega^2}$.  In the experiment, $L = 55~\mu$m, implying
$\mu = 2\pi\hbar\times 15$~Hz.  Equation (\ref{eq:num}) can be normalized
to the total number of atoms $N$ using $\int n_{1D}dy=N$,
which yields $\mu/g_{1D}=3N/4L$.

The interaction potential between the packets is
$U\sub{int}=2g_{1D}n_{1D}$, where the factor of two comes
from the exchange effect; however once split the number density is reduced by a
factor of two, leaving
\begin{equation}
U\sub{int}(\xi_1)=\left\{
\begin{array}{l l}
  \mu\left(1-\displaystyle{\frac{\xi_2^2}{L^2}}\right) & \quad -L<\xi_2<L \\
  0 & \quad \mbox{otherwise}\\ \end{array} \right .  \label{eq:Uint}
\end{equation}
An atom at position $\xi_1$ in the
first packet will experience the potential due to atoms at $\xi_2$ in the
second packet.  For times small compared to $\omega^{-1}$ we relate
two positions by approximating the packet centres as $y_{1,2}=\pm v_0t$,
so that position $\xi_1$ in the first packet coincides
with $\xi_2-2v_0t$ in the other packet.
The phase difference due to interactions is then
\begin{equation}
\phi\sub{int}(\xi)=\frac{1}{\hbar}\int_0^{2\tau_1}[U\sub{int}(\xi-
2v_0t)-U\sub{int}(\xi+2v_0t)]dt .
\label{eq:Pint}\end{equation}
This can be evaluated as
\begin{equation}
\phi\sub{int} =
\left\{ \begin{array}{l l} 8 k\xi (\omega \tau_1)^2 & \mbox{if}~\xi < L-2v\tau_1\\
\rule{0pt}{4ex}
\displaystyle\frac{m\omega^2}{\hbar v_0} \xi(L-\xi)^2
& \mbox{if}~\xi > L-2v\tau_1
\end{array}\right .
\end{equation}
In our experiment, the packets separate at $\tau_1 > L/v_0 = 4.7$~ms.
For times larger than this, the second expression applies at all positions
in the packet.

\begin{figure}
{\includegraphics[width=6.5in]{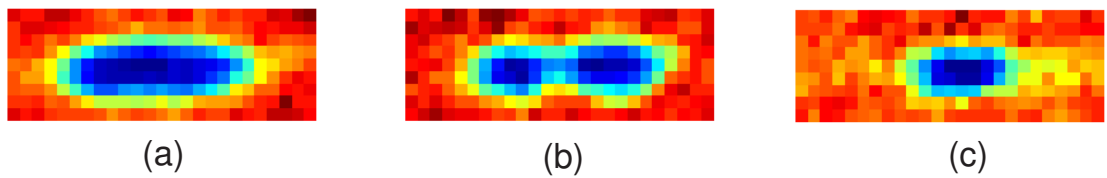}}
\caption{\label{fig:cloud}(colour) Absorption images of wave packets
used in the condensate interferometer. (a) The condensate
distribution, before the interferometer operation. (b) A packet
obtained at the output of the interferometer, showing the effect of
a phase gradient.  (c) Another output packet, showing the effect of
shifting the overall phase of the interferometer by $\pi$.}
\end{figure}

When the packets are finally recombined, the fraction of atoms
brought to rest is $[1+\cos(\theta + \phi)]/2$, where
$\theta$ is an applied phase and $\phi$ is the confinement
phase calculated above.  Since $\phi$ depends on position,
the final packet density will exhibit a sinusoidal modulation.
This can be experimentally observed as follows:
The recombination pulse generally produces
three packets, at velocities of $0, v_0$ and $-v_0$.
After the pulse, we wait some time for these packets to separate and
then use a resonant probe beam to take an absorption image.
Figure~\ref{fig:cloud} shows examples of the images
obtained.  Image (a) shows the initial condensate, with no gradient.
Images (b) and (c) show packets obtained after the interferometer
sequence, with overall phase differing by $\pi$.
Both images show a modulation
of the atomic density, with suppression at either the centre (a) or
edges (b) of the packet.

To analyze such images, we integrated the absorption signal in the direction transverse to the interferometer axis, thereby
obtaining a one-dimensional profile.  We then
simultaneously fit up to 12 packets (four images of three packets each) to a
modulated Thomas-Fermi function with a common gradient $G$
and fixed width $L$:
\begin{equation}
A(y)=A_0+B\sin^2\big[G(y-y_0)+\theta\big]
\times \textrm{min}\!\left[0,\ 1-\left(\frac{y-y_0}{L}
\right)^2\right].
\label{eq:fit}\end{equation}
Each of the images in a set is taken with the same experimental
conditions except for the applied phase $\theta$, which is varied.
From the fit, we acquire $G=\frac{1}{2}|d\phi/d\xi|$.
For gradients comparable to $1/L$ or smaller, the fits become less reliable
as there is little modulation to observe.

\begin{figure}
{\includegraphics[width=3in]{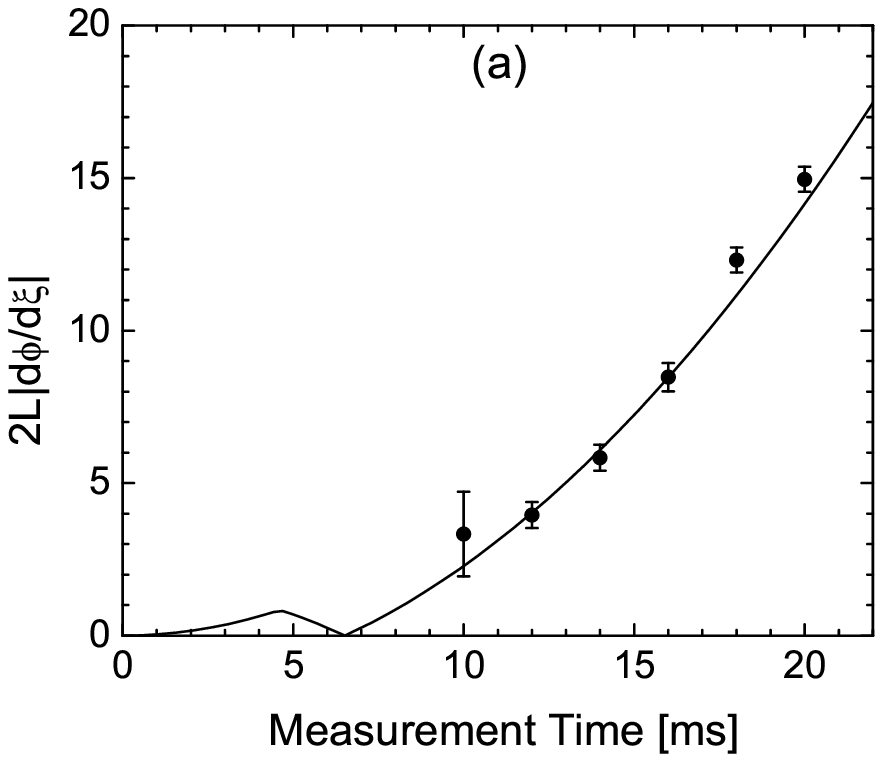}}{%
\includegraphics[width=3in]{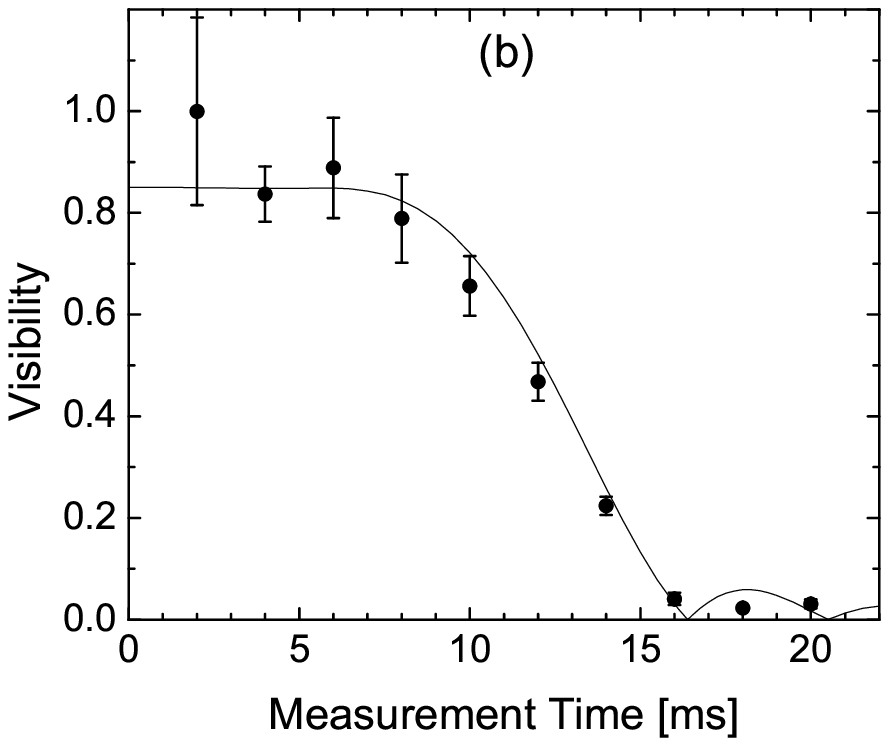}}
\caption{\label{fig:onevis} (a) Magnitude of the phase gradient
for a single-sided interferometer
of total duration $2\tau_1$.  Points show the experimental measurements,
with error bars determined by the fitting procedure.
The curve shows the results of the calculation described in the text,
with the gradient scaled to the length of the packet, $2L$.
(b) Visibility of the single-sided interferometer, as a function of
total duration.  
Points show the measured values, while the curve shows
the theoretical prediction.  
The prediction has been scaled by 0.85 to
account for experimental imperfections that are expected to be
independent of duration.}
\end{figure}

Figure~\ref{fig:onevis}(a) shows the measured and calculated gradients
as functions of the total measurement time $2\tau_1$.  The piecewise
behaviour for small $\tau_1$ is caused by the interaction effect.
The gradient from interactions depends on position in the packet; here
we plot the gradient at the centre $\xi = 0$.
The measured gradients agree well with the theoretical expectation.

Normally, we operate the interferometer by measuring the total
number of atoms in each of the three packets to determine 
$N_0/N$.  This signal can be evaluated as
\begin{equation}
\frac{N_0}{N}=\frac{1}{2}\left[ 1+\int_{-L}^{L}
n_{1D}(\xi)\cos[\theta+\phi(\xi)]d\xi\right].
\label{eq:Nf}\end{equation}
Using the symmetry of $\phi$ this reduces to
\begin{equation}
\frac{N_0}{N}=\frac{1}{2}\left[1+\cos(\theta)\frac{3}{4L}\int_{-L}^{L}
\left(1-\frac{\xi^2}{L^2}\right)\cos\phi\, d\xi \right],
\end{equation}
from which the visibility of the interference is seen to be
\begin{equation}
V=\frac{3}{4L}\int_{-L}^{L}
\left(1-\frac{\xi^2}{L^2}\right)\cos\phi\ d\xi.
\label{eq:V}
\end{equation}
Experimentally, the visibility and its uncertainty
are measured by scanning $\theta$ and
fitting the resulting signal to the form $(1+V\cos\theta)/2$.
Figure~\ref{fig:onevis}(b) shows a comparison of the measured and predicted
visibility curves.  Even at short times, we obtain typical visibilities of
about 0.85, due to imperfections in the standing wave operations and
other experimental effects.  We therefore scale the theoretical curve by
this factor.  Again, good agreement is obtained.

The figure shows that the visibility is reduced by one half at
$2\tau_1 \approx 12$~ms, corresponding to a gradient of 4 radians
across the full packet.
At this time, the
centres of the clouds are 120 $\mu$m apart, meaning the packets are
just separated.  This makes it difficult, for instance, to
apply a field to one packet without influencing the other, and
yields a figure of merit $\chi$ of only
$0.07~\mu$m$\cdot$s.
It is possible to operate the interferometer with larger phase gradients,
by considering the density profile
as above and taking the fit phase $\theta$ in Eq.~\ref{eq:fit}
as the interferometer output.  However, this introduces considerable noise
in practice because of imaging imperfections such as the
limited spatial resolution.

The results of this section illustrate clearly that the operation of the
single-sided interferometer is limited by phase gradient effects.

\section{Double-sided interferometer}

The confinement phase gradient is alleviated in a double-sided
interferometer, where the gradients from the two halves tend to
cancel out.  To analyze this situation, consider an asymmetric interferometer
with leg duration $\tau_1$ in the first half and $\tau_2$ in the second half,
as in Fig.~\ref{fig:Tra}(b).
The calculation of (\ref{eq:phi}) is readily extended
to this case, yielding
\begin{equation}
 \phi = 4k\xi \left\{ \cos\left[2\omega (\tau_1+\tau_2)\right]
-2\cos\left[\omega(\tau_1+2\tau_2)\right]   +2\cos(\omega \tau_2) -1 \right\}.
\label{eq:DPhi}
\end{equation}
In the case $\tau_1 = \tau_2 = \tau$, this gives to leading order a
gradient $d\phi/d\xi = 16k (\omega \tau)^4$, suppressed by a factor
of $(2\omega\tau)^2$ compared to the single-sided interferometer.
The gradient due to interactions can be calculated as well, but here
the maximum gradient obtained for separated arms becomes so small to
be nearly negligible. For both sources of gradient, the cancellation
is not perfect because the packet velocities during the two halves
are not exactly equal, as discussed above. Since the velocity
correction $\delta$ is on the order of $v_0 (\omega\tau)^2$, the
scaling of the residual gradient is sensible.

\begin{figure}
{\includegraphics[width=3in]{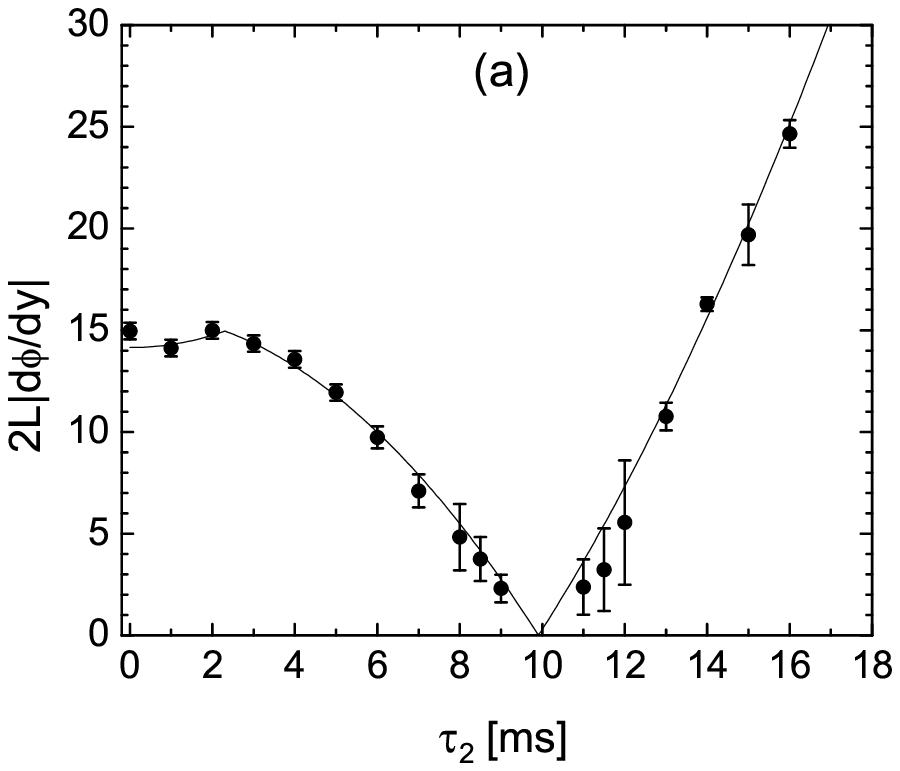}}%
{\includegraphics[width=3in]{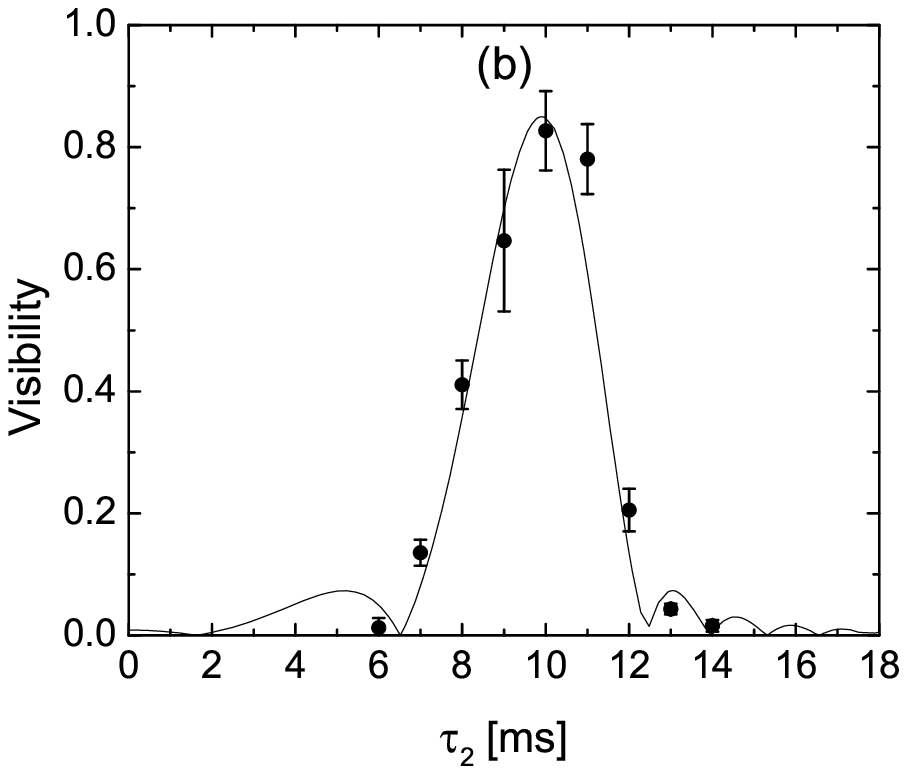}}
\caption{\label{fig:grad} (a) Phase gradient observed in an
asymmetric double-sided interferometer. The first half of the
trajectory used leg time $\tau_1$ and the second half $\tau_2$.
Here $\tau_1 = 10$~ms and $\tau_2$ is varied. When $\tau_2 =
\tau_1$, the phase gradient is nearly eliminated.  (b)
Interferometer visibility as $\tau_2$ is varied.  For both plots,
data points show the experimental results and the curves show the
model calculation. In (b), the calculation is scaled by 0.85 as in
Fig.~\protect\ref{fig:onevis}.}
\end{figure}

We observe the cancellation effect by fixing $\tau_1 = 10$~ms and
varying $\tau_2$.  The results are seen in Fig.~\ref{fig:grad}(a),
where the gradient becomes unobservably small when $\tau_2 =
\tau_1$. The effect of interactions is seen at
$\tau_2 < 2$~ms. Figure \ref{fig:grad}(b) shows how the resulting
visibility is recovered. We note that the cancellation effect is
very much analogous to the spin echo phenomenon known in magnetic
resonance studies.

\begin{figure}
{\includegraphics[width=3in]{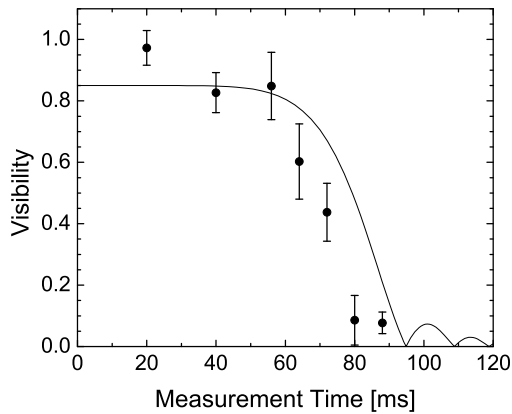}}
\caption{\label{fig:2vis}
Visibility of symmetric double-sided interferometer, as a function of
the total measurement time $T = 4\tau$.  Data points show the
experimental results and the curve shows the model prediction, scaled
by 0.85 as in Fig.~\protect\ref{fig:onevis}.}
\end{figure}

Using this approach, the measurement time of the interferometer can be
significantly extended, as seen in Fig.~\ref{fig:2vis}.
The data agree reasonably well with the predicted values,
but in this case some caution in the interpretation is warranted.
At such long times, the approximation that the internal dynamics of the
packet is negligible is no longer well justified.  The period of oscillation
in the tightest direction of the guide is only 166~ms, and interactions
will couple motions in the transverse and longitudinal directions.
Because of this, the length of the condensate can be expected to vary on
these time scales, which will alter the phase gradient.  Given that
the predicted gradient is already the difference of two relatively
large values, even a small effect from the internal dynamics could
be significant.  Accurate calculation of the packet dynamics would require the 
numerical solution of the full three-dimensional Gross-Pitaevskii equation.

Experimental evidence of the inadequacy of the simple theory presented
in Section 2 is that as the visibility of the double-sided interferometer
falls to zero, images of the atoms do not exhibit clear gradients like those
seen in Fig.~\ref{fig:cloud}.  Instead, the recombination appears to be
uniform.  A possible explanation is that transverse gradients develop
along with the longitudinal ones, since the absorption imaging
technique averages over phase variations that are parallel to the probe beam.
Visibility will also be lost if the
standing-wave laser is misaligned, so that the packets are not
spatially overlapped when the recombination pulse is applied.
We ruled this out by observing the packets along two axes
to verify their alignment in all three dimensions.  A third possibility
is overall phase noise such as that resulting from vibrations of the
mirror producing the standing wave.  This would cause the interferometer
signal to differ from shot to shot and reduce the visibility upon
averaging.  However, we do not observe the shot-to-shot variations that
would characterize such noise.  For instance, at at total
measurement time $T = 100$~ms, the deviation in $N_0/N$ from one
experiment to the next is only 1\%.

These experiments demonstrate that the double-sided interferometer extends
the useful measurement time of the interferometer by a factor of
six compared to the single-sided case.  This improves the arm separation
by a factor of three, allowing the packets to be well separated and
individually accessible.  
We note that a double-sided interferometer is not suited for
some types of measurements.  In a measurement of gravity, for instance,
the gravitational phase of the two arms would cancel just as the confinement
phase does.
However, many other measurements remain possible, including the
Sagnac phase in a loop geometry and any case where
the effect to be measured can be applied at a definite time.
We have recently used this technique to interferometrically measure the
dynamic polarizability of $^{87}$Rb \cite{Deissler07}.

\section{Free-oscillation interferometer}

Unfortunately, the same trick cannot be repeated to extend
the arm separation even further.  We theoretically considered
a `four-sided' interferometer, where the packets pass through
each other three times before being recombined.  This did
reduce the phase gradient, but only by a factor of four, rather than
by a power of $\omega\tau$.  Given the $\tau^4$ scaling of the gradient,
this would provide only a modest increase in arm separation.  At
the same time, imperfections in the reflection operations begin
to have a significant effect, so we did not pursue this in
the experiment.

Instead, we implemented an
interferometer in which no reflection pulses are used.
The packets are split, allowed to complete a full natural oscillation
in the guide, and then recombined as shown in Fig. \ref{fig:Tra}(c).  Insofar as the
guide potential is harmonic, the trajectories for the
two half-cycles will be symmetric and any phase gradients should
cancel precisely.  However, the oscillation period in our guide
is 0.9 s.  Over such a long time scale, many external noise sources
can impart an uncontrolled phase difference to the packets, making
the interferometer output effectively random.  We were therefore
unable to observe a controlled interference signal as in the previous
experiments, but rather used the run-to-run variations in the output
as a measure of the interference visibility.  This technique, including
the use of a freely oscillating trajectory, was also demonstrated
by Segal {\em et al.} \cite{Segal07}.  Their results corresponded to
a figure of merit $\chi = 40~\mu$m$\cdot$s.

\begin{figure}
{\includegraphics[width=3in]{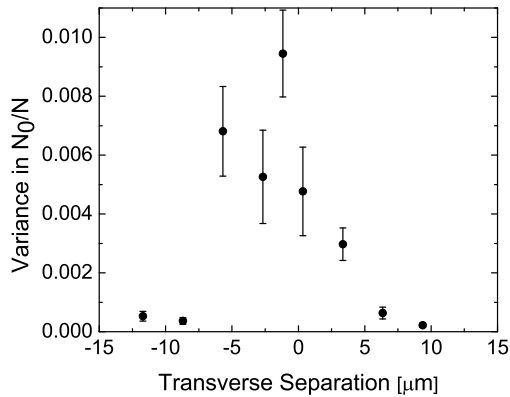}}
\caption{\label{fig:onesec}
Interference for atoms freely oscillating
in the waveguide.  The measurement time is 0.9~s, corresponding to a separation
of 1.7 mm.  The phase of the interference is random, so the variance of
the signal measured over several experiments is plotted against
the transverse separation of the wave packets at the time of the
recombination pulse.  The large increase in variance when the packets
are overlapped is taken as a signal that interference is present.
The error bars represent the error in the variance, estimated as
$\Delta^2/N$ for variance $\Delta^2$ and number of measurements $N$.}
\end{figure}

For such long measurement times, the alignment of the
standing wave beam to the guide axis is critical, to avoid
having the packets miss each other in the transverse directions.
In Fig.~\ref{fig:onesec}, we show the variance of $N_0/N$ observed as
the beam alignment angle is varied, plotted against the measured
packet displacement.  The clear peak indicates that interference
is present.  Assuming the fluctuations represent an underlying
distribution $(1 + V\cos\theta)/2$ with random $\theta$, the variance
$\Delta^2$ is related to the visibility $V$ by $\Delta^2 = V^2/8$,
thus suggesting $V = 0.3$.  A possible reason for the imperfect
visibility is the effect of internal dynamics introducing asymmetry
to the two halves of the trajectory.
We note that no interference
was observed when the atoms were allowed to oscillate for only one half
of the axial period.

In this experiment, the maximum arm separation is 1.7~mm.  Without
control of the overall phase, it is not possible to make use of this
arm separation in a practical measurement, but it can be hoped that
this is a technical problem that is solvable through various
stabilization methods.  If so, this type of interferometer would
offer an unprecedented arm separation and measurement time.  

\section{Conclusions/Acknowledgments}

We have studied the effect of longitudinal confinement on a guided wave
interferometer, under various conditions.  It is seen conclusively that for a
single-sided interferometer, large phase gradients induced by the
confinement potential place a severe limit on the usable arm separation of the
the interferometer.  The phase gradients were measured and found to
agree well with a simple theory in which the internal dynamics of
the atomic wave packets are neglected.

The interferometer performance can be significantly improved using a
double-sided trajectory, in which each packet experiences each side
of the potential.  In this case, measurement times of up to 72~ms
and arm separations of 420~$\mu$m are achieved.  An even larger arm
separation of 1.7~mm can be obtained by allowing the packets to
complete a full oscillation in the guide, with no imposed reflection
pulses.  In this case, however, noise effects introduce a random
phase shift that prevents the device from being practically useful.
When the measurement time of the interferometer becomes comparable
to the motional period in the transverse directions of the guide,
accurate predictions of performance will likely require a more
sophisticated model in which the packet dynamics are included.

One conclusion to be drawn is that axial confinement in a
guided wave interferometer is generally a detriment and should be avoided.
In a linear interferometer, the axial potential typically comes from the
electrical connections at the end of the guide, so simply making
the guide longer would improve the flatness \cite{Wu05}.  
Alternatively, additional
current elements could be added to raise the potential in the centre
of the guide.  It is also worth noting that these effects should
generally be suppressed in a ring-shaped guide.  Even if the potential
around the ring is not perfectly uniform, the net phase gradient will
be close to zero if the packets are
allowed to propagate around the full circumference.
In this case no reflection pulses are required, making it similar to
the free-oscillation interferometer demonstrated here.
Through considerations such as these, we expect that
the findings here will be useful for guiding future designs.

We are grateful to J. Stickney and A. Zozulya for useful conversations
on these subjects.  This work was funded by the U.S. Defense Advanced
Research Projects Agency, under grant number W911NF-06-1-0474.

\end{document}